\documentclass[%
 aip,
 amsmath,amssymb,
 reprint,%
]{revtex4-1}

\usepackage{float}
\usepackage{graphicx}% Include figure files
\usepackage{caption}
\usepackage{ragged2e}
\usepackage{dcolumn}% Align table columns on decimal point
\usepackage{bm}% bold math
\usepackage[utf8]{inputenc}
\usepackage[T1]{fontenc}
\usepackage{mathptmx}
\usepackage{etoolbox}
\usepackage{subcaption}
\usepackage{adjustbox}
\makeatletter
\def\@email#1#2{%
 \endgroup
 \patchcmd{\titleblock@produce}
  {\frontmatter@RRAPformat}
  {\frontmatter@RRAPformat{\produce@RRAP{*#1\href{mailto:#2}{#2}}}\frontmatter@RRAPformat}
  {}{}
}%
\makeatother
\begin{document}

\preprint{AIP/123-QED}

\title{Undetected Photon Spectroscopy in the Long-wave and Far Infrared
}
%\title{Long-wave infrared undetected photon spectroscopy}
% Force line breaks with \\
\author{Jinghan~Dong}
 \email{jinghandong9@gmail.com}

\author{Jingrui~Zhang}%
\author{Haichen~Zhou}%
\author{Weijie~Nie}%
\author{John~G.~Rarity}%
\author{Alex~S.~Clark}
\affiliation{ 
Quantum Engineering Technology Labs, H.H. Wills Physics Laboratory and School of Electrical, Electronic and Mechanical Engineering, University of Bristol, BS8 1TL, Bristol, UK}
\date{\today}% It is always \today, today,

\begin{abstract}
Spectroscopy with undetected photons allows the spectral response of a sample at long wavelengths to be measured through interference of a shorter-wavelength signal field, without directly detecting the light that probes the sample. Previous demonstrations of undetected-photon spectroscopy in the molecular fingerprint region have reached approximately \(10.5~\mu\mathrm{m}\), leaving longer wavelengths largely unexplored. Here, we extend this approach to \(20~\mu\mathrm{m}\) wavelength, demonstrating infrared undetected-photon spectroscopy using a BaGa\(_2\)GeSe\(_6\) (BGGSe)-based nonlinear interferometer while only the near-infrared (NIR) signal is detected using a silicon detector. Rotation of the BGGSe crystal provides broad wavelength tuning, and Fourier-transform analysis of the signal interferograms yields the corresponding idler spectra and sample transmittance. Broadband absorption spectroscopy is demonstrated using Kapton across a \(7.8\text{--}20~\mu\mathrm{m}\) wavelength range, with the measured absorption agreeing with independent infrared measurements. The same system is further employed to detect sulfur hexafluoride (\(\mathrm{SF_6}\)) gas via spectroscopy near \(10.55~\mu\mathrm{m}\). These results substantially extend the accessible spectral range of spectroscopy with undetected photons while retaining silicon-based NIR detection.
\end{abstract}

\maketitle

\section{Introduction}

Infrared spectroscopy is widely used for chemical identification because many molecules have fundamental vibrational absorption bands in the mid-infrared that are stronger and more selective than the corresponding overtone bands in the near-infrared (NIR)~\cite{stuart2004infrared,haas2016advances}. The molecular fingerprint region is often defined as \(1500\text{--}500~\mathrm{cm}^{-1}\), corresponding to \(6.7\text{--}20~\mu\mathrm{m}\) wavelength range~\cite{timmers2018molecular}. In this region, highly molecule-specific absorption features provide strong spectral signatures for chemical identification and sensing.
However, operating in this wavelength range can be difficult due to the requirement for specialised infrared hardware. Fourier-transform infrared spectroscopy provides broadband spectral measurements, while laser-based techniques using quantum-cascade lasers, optical parametric oscillators, and difference-frequency generation sources can probe selected absorption bands with high sensitivity~\cite{griffiths1983fourier,stuart2004infrared,vainio2016mid}. Broadband measurements can also be performed using supercontinuum sources and frequency-comb spectroscopy~\cite{zorin2022advances,timmers2018molecular}. Despite their different approaches, spectroscopy at longer infrared wavelengths generally requires sources, optics, and detectors suitable for the target spectral range~\cite{griffiths1983fourier,stuart2004infrared,haas2016advances,zorin2022advances}. In particular, high-performance photon detectors at long infrared wavelengths often require cooling for low-noise operation~\cite{rogalski2023infrared}, although uncooled thermal detectors are also available~\cite{piotrowski2004uncooled}. By contrast, silicon detectors offer low noise, high sensitivity, and comparatively low cost in the visible and NIR. This motivates spectroscopic approaches that access longer-wavelength molecular signatures while retaining detection in the NIR.

Nonlinear interferometry provides such a route. In a nonlinear interferometer, signal--idler photon pairs are generated in a nonlinear crystal through spontaneous parametric down-conversion. The idler can interact with a sample at a longer wavelength, while the signal is detected in the NIR. When the two possible pair-generation alternatives are made indistinguishable, loss and phase changes in the idler path modify the interference of the detected signal~\cite{PhysRevA.44.4614}. In this way, infrared absorption information can be recovered without directly detecting the infrared photons. This principle has been used for imaging, microscopy, optical coherence tomography, Fourier-transform spectroscopy, and gas sensing with undetected photons~\cite{pearce2023practical,gemmell2023loss,lemos2014quantum,kviatkovsky2020microscopy,vanselow2020frequency,hashimoto2024fourier,dong2025methane,dong2025open,lindner2021nonlinear,neves2024open}.

Recent work has demonstrated several approaches to spectroscopy with undetected photons, including Fourier-transform spectroscopy~\cite{mukai2022quantum,hashimoto2024fourier}, broadband spectroscopy~\cite{paterova2022broadband,tashima2024ultra}, wavelength-tunable absorption spectroscopy~\cite{arahata2022wavelength}, and related nonlinear-interferometric sensing schemes~\cite{kumar2025controlling}. These developments have improved the spectral coverage, resolution, and sensitivity of nonlinear-interferometric spectroscopy. However, experimental studies have remained concentrated at shorter infrared wavelengths. In the molecular fingerprint region, spectroscopy with undetected photons has been demonstrated at wavelengths up to \(10.5~\mu\mathrm{m}\)~\cite{arahata2022wavelength,paterova2022broadband,mukai2022quantum}, but experimental demonstrations at longer wavelengths remain limited. Extending the accessible spectral range would provide access to additional molecular absorption features while retaining shorter-wavelength detection.

\begin{figure*}[t]
    \centering
    \includegraphics[width=\textwidth]{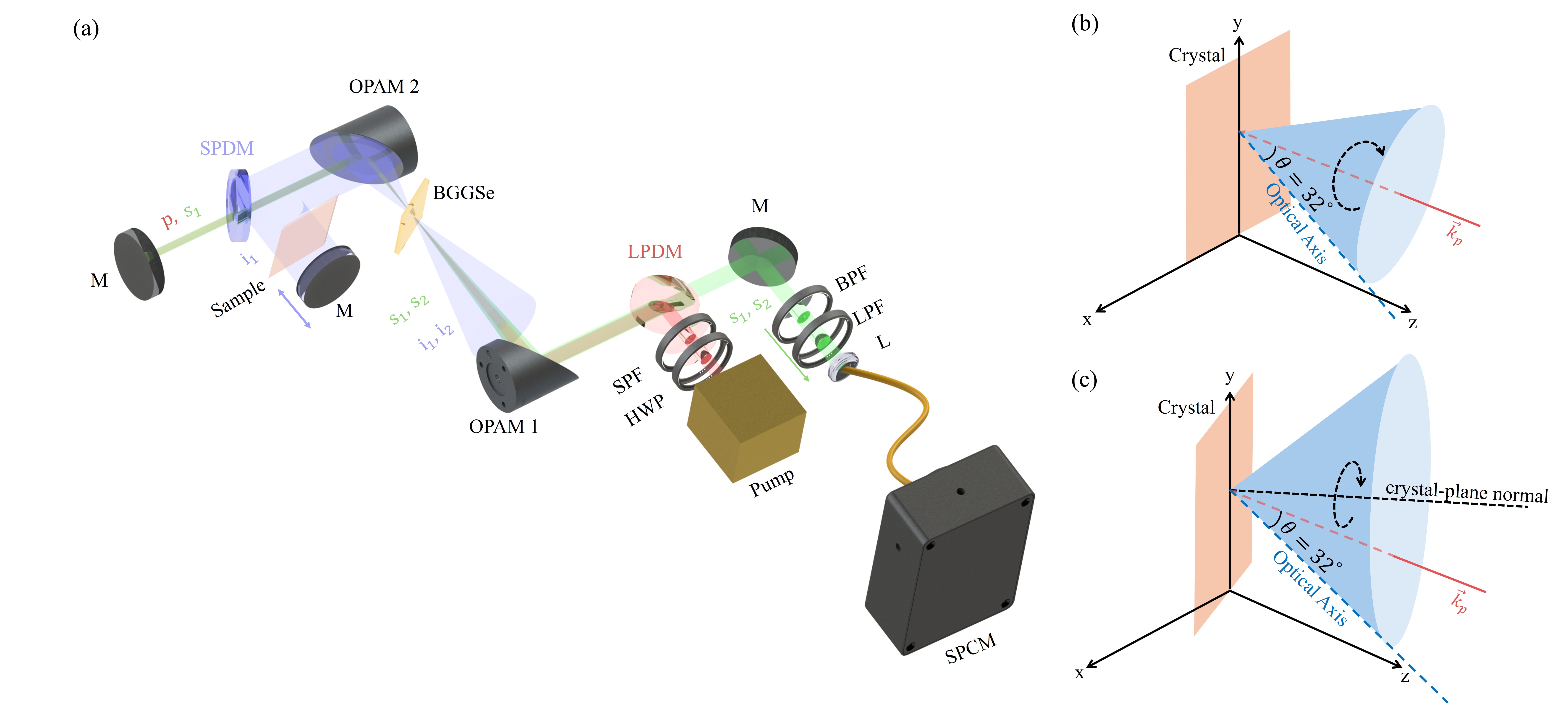}
   \caption{\justifying
Experimental configuration and crystal-angle tuning geometry. 
(a) Michelson-type nonlinear interferometer used for undetected-photon spectroscopy from \(7.8\) to \(20~\mu\mathrm{m}\). A \(785~\mathrm{nm}\) pump is focused into the BGGSe crystal, generating first-pass signal and idler probability amplitude (field) in modes labelled \(s_1\) and \(i_1\). The idler is reflected by the SPDM and passes through the sample twice, while the pump and signal propagate through the other interferometer arm. The pump, \(s_1\), and \(i_1\) return to the BGGSe crystal, where a second probability amplitude for signal and idler generation in modes \(s_2\), and \(i_2\) is generated. When first pass and second pass modes overlap the probability amplitudes add leading to interference effects in the rate of pair photon generation which can be seen in singles rates in both signal and idler arms. This allows measurement of phase and amplitude modulation in the idler arm using only the signal photons which are measured using a silicon SPCM.
(b) Ideal normal-incidence geometry, in which the crystal surface normal is parallel to the pump wave vector \(\mathbf{k}_p\). Rotation about the surface normal leaves the angle between \(\mathbf{k}_p\) and the crystal optical axis nearly unchanged.
(c) Experimental oblique-incidence geometry. After applying a fixed crystal tilt, rotation about the crystal-plane normal changes the orientation of the optical axis relative to \(\mathbf{k}_p\), thereby tuning the phase-matching condition.
M: mirror, HWP: half-wave plate, SPF: short-pass filter, LPF: long-pass filter, BPF: band-pass filter, LPDM: long-pass dichroic mirror, SPDM: short-pass dichroic mirror, OAPM: off-axis parabolic mirror, L: lens, SPCM: single-photon counting module.}
    \label{fig:setup}
\end{figure*}

In this work, we develop a nonlinear interferometer that uses a BaGa\(_2\)GeSe\(_6\) (BGGSe) nonlinear crystal to extend spectroscopy with undetected photons to longer wavelengths within the molecular fingerprint region, spanning \(7.8\text{--}20~\mu\mathrm{m}\). BGGSe was selected as the nonlinear crystal because it has a broad transmission band from \(0.58~\mu\mathrm{m}\text{--}18~\mu\mathrm{m}\)~\cite{badikov2016crystal,petrov2021barium} and a large second-order nonlinear coefficient~\cite{kato2020effective}. BGGSe has also been studied and used for nonlinear frequency conversion in the mid-infrared and long-wave infrared~\cite{grechin2020baga2ges6,elu2020few,chen2024narrowband,guo20255}. These properties make BGGSe a promising nonlinear material for extending undetected-photon spectroscopy to substantially longer wavelengths. The idler wavelength is tuned by rotating the single BGGSe crystal, while the corresponding NIR signal is measured using a silicon detector. Fourier-transform spectroscopy is used to obtain wavelength-dependent information from the undetected idler. Broadband spectroscopy of Kapton tape was performed to evaluate the spectral capability of the system, and was in good agreement with measurements from a commercial FTIR spectrometer. Finally, we perform gas sensing of sulfur hexafluoride (\(\mathrm{SF_6}\)) which is widely used in high-voltage electrical equipment but is also a potent greenhouse gas with a long atmospheric lifetime~\cite{maiss1998atmospheric,cui2024recent}. Its strong absorption band near \(10.55~\mu\mathrm{m}\) provided a suitable test of molecular gas spectroscopy using this nonlinear interferometer~\cite{gordon2022hitran2020}, with the results matching those taken with a commercial FTIR system and with the HITRAN spectral database.

\section{Results}

\subsection{Nonlinear interferometer setup}

We use the Michelson-type nonlinear interferometer arrangement\cite{Herzog1994}  adapted here to generate widely non-degenerate photon pairs to perform undetected-photon spectroscopy. A fibre-coupled 785 nm continuous-wave pump was focused into a 2 mm-thick BGGSe crystal. In the first pass through the crystal there is a probability amplitude for a pair photon to be created in signal \(s_1\) and idler \(i_1\) modes. The idler field passes through the sample twice and returns to the crystal while the pump and signal are reflected back via the other interferometer arm. On the second pass through the crystal, a second probability amplitude for generation of a pair in modes \(i_2\), \(s_2\) is created. When first pass and second pass modes overlap we see interference effects leading to enhancement or suppression of photon pair generation dependent on any added phase in the 
idler arm which can be detected in 
both the idler and signal output modes \(i_2\) and \(s_2\). However, we only need to measure the signal \(s_2\) in the near infrared silicon single-photon counting module (SPCM) to see phase and fringe visibility variations produced by our sample in the idler arm. Further details of the optical setup and components are provided in the Materials and Methods section.

%Figure~\ref{fig:setup}(a) shows the Michelson-type nonlinear interferometer used for undetected-photon spectroscopy. A fibre-coupled \(785~\mathrm{nm}\) continuous-wave pump was focused into a \(2~\mathrm{mm}\)-thick BGGSe crystal. In the first pass through the crystal, the signal \(s_1\) and idler \(i_1\) were generated. The idler passed through the sample twice and returned to the crystal. The pump and signal propagated through the other interferometer arm. On their return to the crystal, a second signal field \(s_2\) was generated. Induced coherence from \(i_1\) resulted in interference between \(s_1\) and \(s_2\), which was measured using a silicon single-photon counting module (SPCM). Further details of the optical setup and components are provided in the Materials and Methods section.

For Fourier-transform spectroscopy, the mirror in the idler arm was scanned while the signal photon counts were recorded as a function of optical path difference. Fourier transformation of the resulting interferogram recovered the spectrum of the undetected idler. Measurements with and without a sample were compared to obtain the sample transmittance. Details of the interferogram acquisition and Fourier-transform analysis are described in the Materials and Methods section.

\subsection{Broadly tunable idler generation}

The phase-matching tuning geometry is shown in Fig.~\ref{fig:setup}(b,c). In the ideal design, the pump is normally incident on the crystal surface, as shown in Fig.~\ref{fig:setup}(b), where the crystal surface normal is parallel to the pump wave vector \(\mathbf{k}_p\). In this configuration, rotation about the surface normal does not change the angle between \(\mathbf{k}_p\) and the optical axis and therefore leaves the phase-matching condition nearly unchanged.

Experimentally, the strongest signal generation was obtained after tilting the crystal away from normal incidence. Once this fixed tilt was established, the crystal was rotated about its surface normal to tune the phase-matching condition, as illustrated in Fig.~\ref{fig:setup}(c). Under this oblique-incidence geometry, rotation changes the orientation of the optical axis relative to \(\mathbf{k}_p\), thereby changing the effective refractive indices and tuning the generated signal and idler wavelengths. Rotation about the surface normal also provided smoother and more reproducible tuning than direct adjustment of the crystal tilt while largely maintaining the interferometer alignment.

This surface-normal rotation angle is denoted by \(\rho\). The position \(\rho=0^\circ\) was defined as the angle giving the strongest generation when using the \(850\text{--}10~\mathrm{nm}\) band-pass filter. Looking along the incoming direction of \(\mathbf{k}_p\), counter-clockwise rotation about the crystal surface normal is defined as positive and clockwise rotation as negative. Sufficient signal generation was maintained across the measured range of \(\rho\), allowing nonlinear interference measurements to be performed throughout the tuning range.

\begin{figure}[t]
    \centering
    \includegraphics[width=\linewidth]{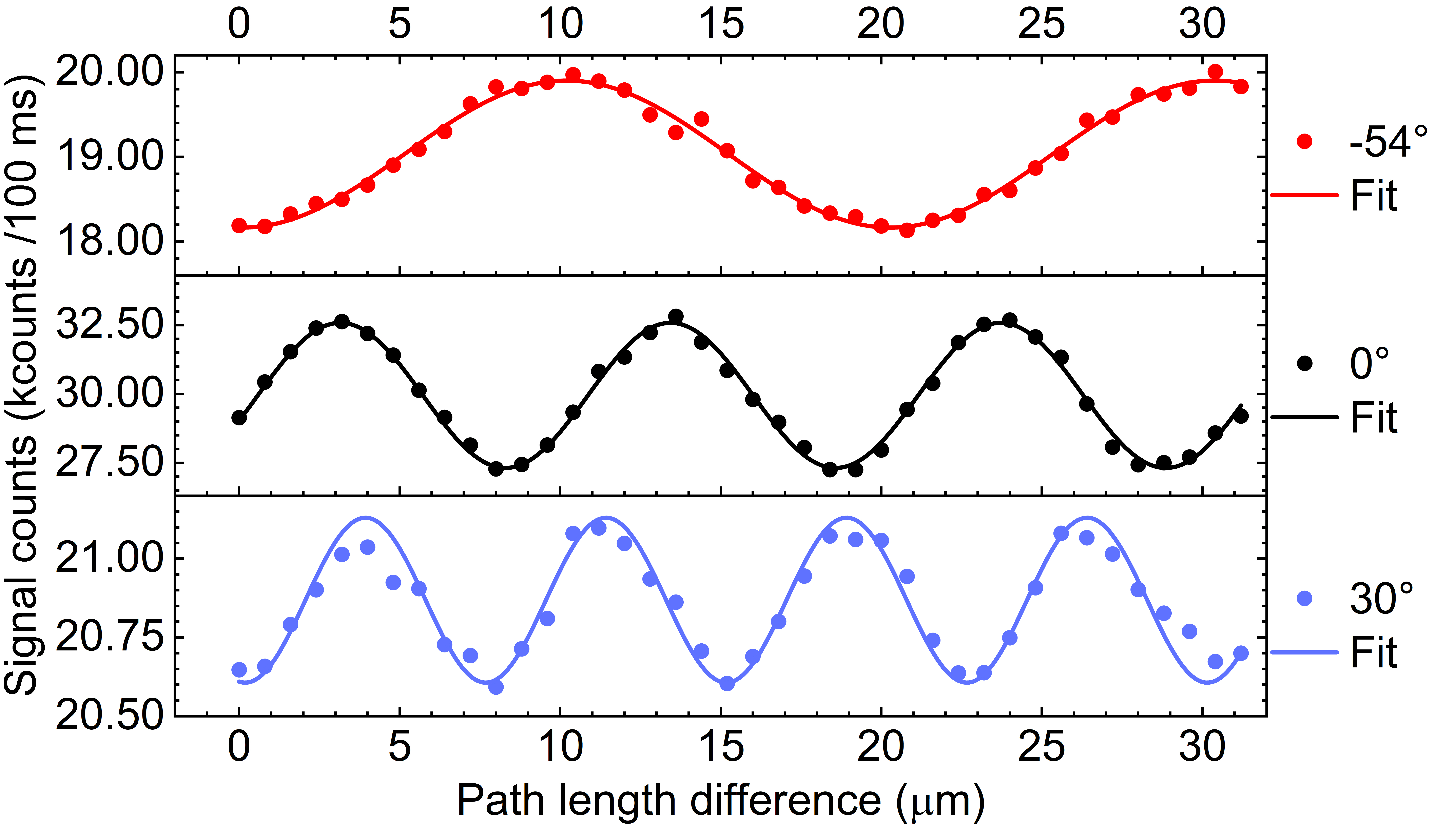}
    \caption{\justifying Representative nonlinear interference fringes measured at \(\rho=-54^\circ\), \(0^\circ\), and \(30^\circ\). The solid curves are sinusoidal fits to the measured signal photon counts.}
    \label{fig:fringes}
\end{figure}

Representative nonlinear interference fringes recorded at three crystal rotation angles are shown in Fig.~\ref{fig:fringes}. At \(\rho=-54^\circ\), the slowly varying fringe gives a fitted period corresponding to an idler wavelength of \(20.19~\mu\mathrm{m}\), with a visibility of \(6.7\%\). At \(\rho=0^\circ\), the shorter fringe period corresponds to \(10.30~\mu\mathrm{m}\), with a visibility of \(11.02\%\). At \(\rho=30^\circ\), a further reduction in fringe period corresponds to \(7.49~\mu\mathrm{m}\), with a visibility of \(1.7\%\). The systematic decrease in fringe period with increasing \(\rho\) is consistent with tuning of the undetected idler towards shorter wavelengths.

These fitted wavelengths represent the dominant fringe periods within the short local scans shown in Fig.~\ref{fig:fringes}, rather than precise measurements of the full idler spectral centers. Since the generated idler has a finite spectral bandwidth, accurate spectral characterization requires interferograms recorded over a substantially longer optical path difference. The idler spectra and their centre wavelengths were therefore determined from longer scans using Fourier-transform analysis.

\begin{figure*}[t]
    \centering
    \includegraphics[width=\linewidth]{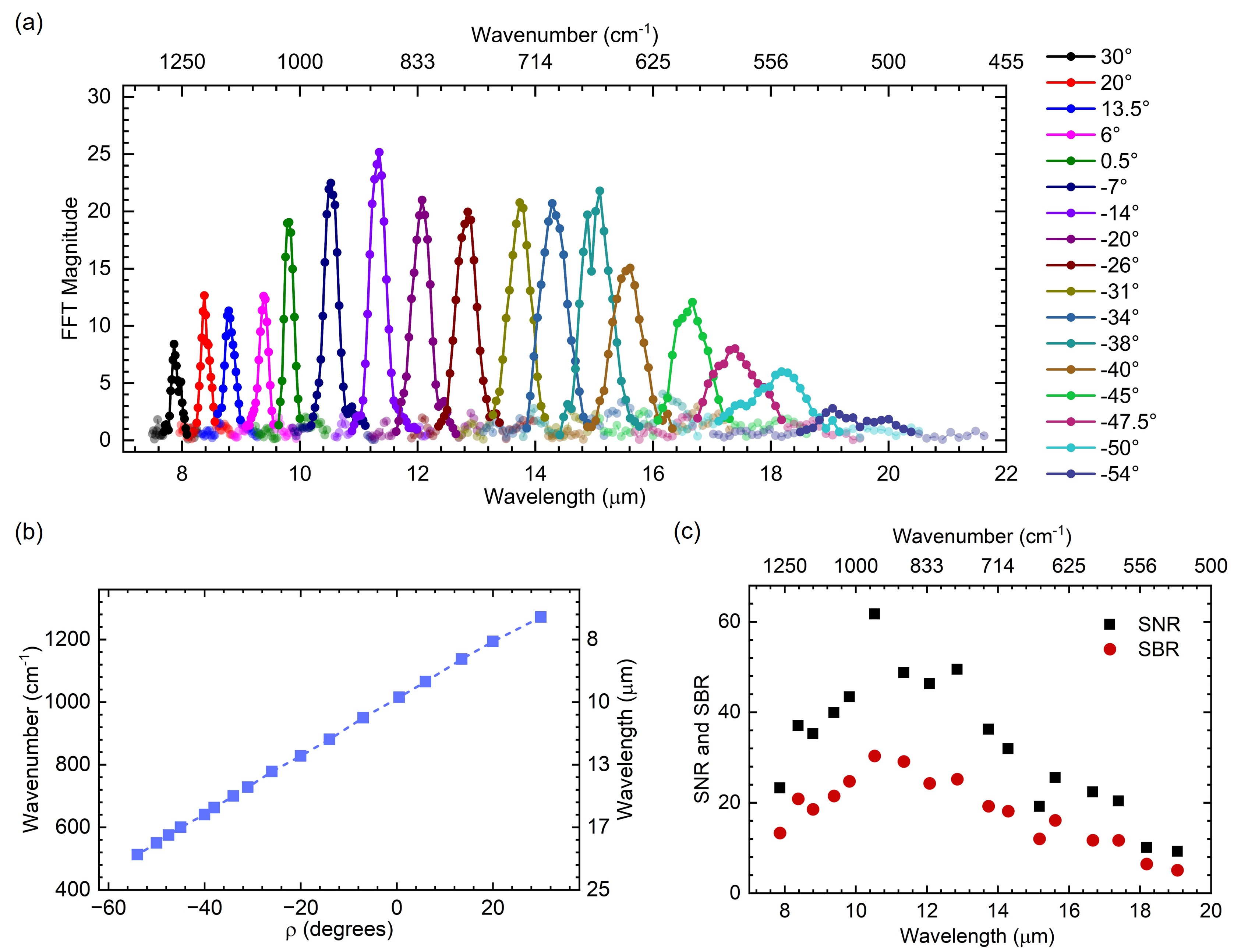}
    \caption{\justifying Crystal-angle tuning and broadband idler generation. 
    (a) Fourier-transform spectra obtained from nonlinear interferograms at different crystal rotation angles \(\rho\). 
    (b) Measured centre wavenumber of the idler spectrum as a function of \(\rho\), with the corresponding centre wavelength shown on the right axis. 
    (c) Signal-to-noise ratio (SNR, black squares) and signal-to-background ratio (SBR, red circles) of the idler spectral peaks as a function of idler centre wavelength. The corresponding wavenumber is shown on the upper axis.}
    \label{fig:angle_spectra}
\end{figure*}

Figure~\ref{fig:angle_spectra}(a) shows the Fourier spectra obtained at different values of \(\rho\). Distinct idler spectral peaks are observed across the full tuning range from approximately \(7.8~\mu\mathrm{m}\) to \(20~\mu\mathrm{m}\). As the crystal is rotated from \(\rho=30^\circ\) to \(-54^\circ\), the idler spectrum shifts progressively towards longer wavelengths, demonstrating broad wavelength tuning with a single BGGSe crystal. The spectral magnitude varies across the tuning range, with the strongest peaks observed in the intermediate wavelength region and reduced amplitudes found towards both the short- and long-wavelength edges. Despite these variations, the idler peaks remain clearly distinguishable from the surrounding background across the full measured range.

The center wavenumber extracted from each spectrum is shown in Fig.~\ref{fig:angle_spectra}(b), showing a monotonic increase that is  approximately linear with the crystal rotation angle over the measured range, providing a simple experimental relation between \(\rho\) and the selected idler spectral region. The accessible range includes the strong \(\mathrm{SF_6}\) absorption band near \(10.55~\mu\mathrm{m}\), and several Kapton absorption bands used in the spectroscopy measurements below.

The spectral quality of the recovered idler peaks was quantified using the signal-to-noise ratio (SNR) and signal-to-background ratio (SBR), as shown in Fig.~\ref{fig:angle_spectra}(c). Both quantities vary with wavelength, with the highest values observed in the intermediate spectral region and lower values towards the edges of the tuning range. The SNR characterizes the peak height relative to fluctuations in the surrounding background, whereas the SBR compares the peak magnitude directly with the mean background level, providing a complementary measure of how clearly each idler peak is separated from the spectral background. Across the full tuning range, the SNR remains above \(9.26\) and the SBR remains above \(5.08\), including for the longest-wavelength spectra approaching \(20~\mu\mathrm{m}\). Together with the systematic shift of the spectral peaks with \(\rho\), these values support the identification of the measured features as tuned idler spectra rather than background fluctuations. The definitions and calculation procedures for the SNR and SBR are given in the Materials and Methods section.

\begin{figure*}[t]
    \centering
    \includegraphics[width=0.9\linewidth]{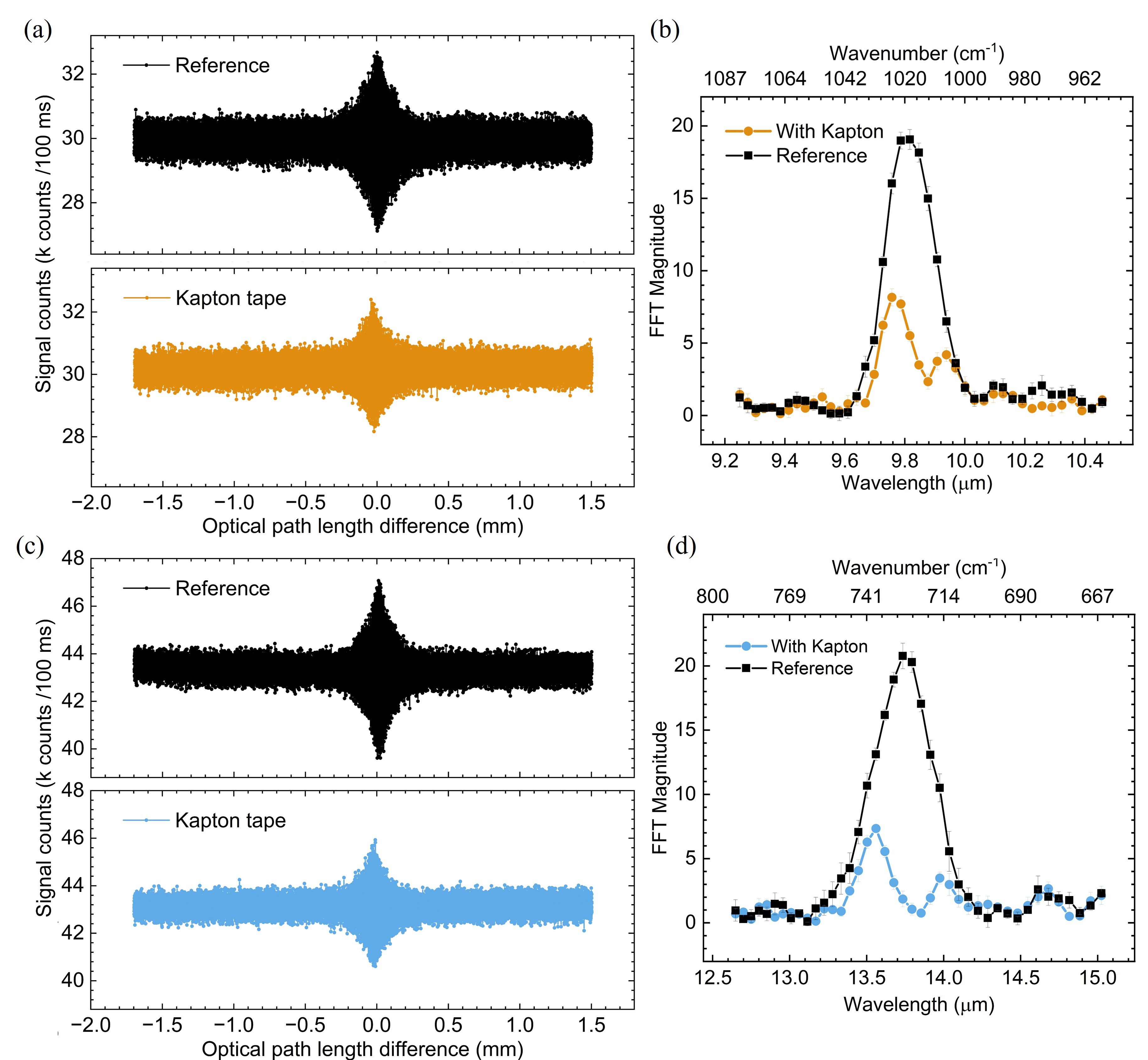}
    \caption{\justifying Interferograms measured with and without a Kapton tape sample in the idler path at (a) \(\rho=0.5^\circ\) and (c) \(\rho=-31^\circ\). The corresponding FFT spectra are shown in (b) and (d), with center wavelengths of \(9.82~\mu\mathrm{m}\) and \(13.73~\mu\mathrm{m}\), respectively.}
    \label{fig:kapton_interferogram}
\end{figure*}

\subsection{Broadband absorption spectroscopy of Kapton}

The spectroscopy capability of the nonlinear interferometer was first tested using Kapton tape with the adhesive removed. Kapton is a high-performance polyimide film adhesive tape that has several strong absorption features across the \(7.8\text{--}20~\mu\mathrm{m}\) range, making it a suitable sample for evaluating the broadband spectroscopic performance of the system.

For each crystal angle, a reference interferogram was first recorded without the sample in the idler path. The Kapton sample was then inserted into the idler arm, and the measurement was repeated under the same conditions. Typical pairs of interferograms are shown in Fig.~\ref{fig:kapton_interferogram}(a) and Fig.~\ref{fig:kapton_interferogram}(c). The interference amplitude is reduced after inserting Kapton because idler absorption unbalances the amplitudes of the first pass and second pass signal-generation. A small shift in the interferogram centre is also observed due to the additional optical thickness introduced by the sample.

The Fourier spectra for two representative values of \(\rho\) are shown in Fig.~\ref{fig:kapton_interferogram}(b) and Fig.~\ref{fig:kapton_interferogram}(d). The black points show the reference spectra without the sample, while the coloured points show the spectra measured with Kapton. The two examples are centred near \(9.82~\mu\mathrm{m}\) and \(13.73~\mu\mathrm{m}\) wavelengths, respectively. In both cases, the FFT magnitude is reduced in the spectral regions where Kapton absorbs the idler. These measurements directly show how wavelength-dependent absorption of the undetected idler modifies the detected signal interferogram and is recovered through Fourier-transform analysis.

The Kapton transmittance was obtained from the ratio of the Fourier amplitudes measured with and without the sample at the same idler wavenumber, as described in the Materials and Methods section.

\begin{figure*}[t!]
    \centering
    \includegraphics[width=\linewidth]{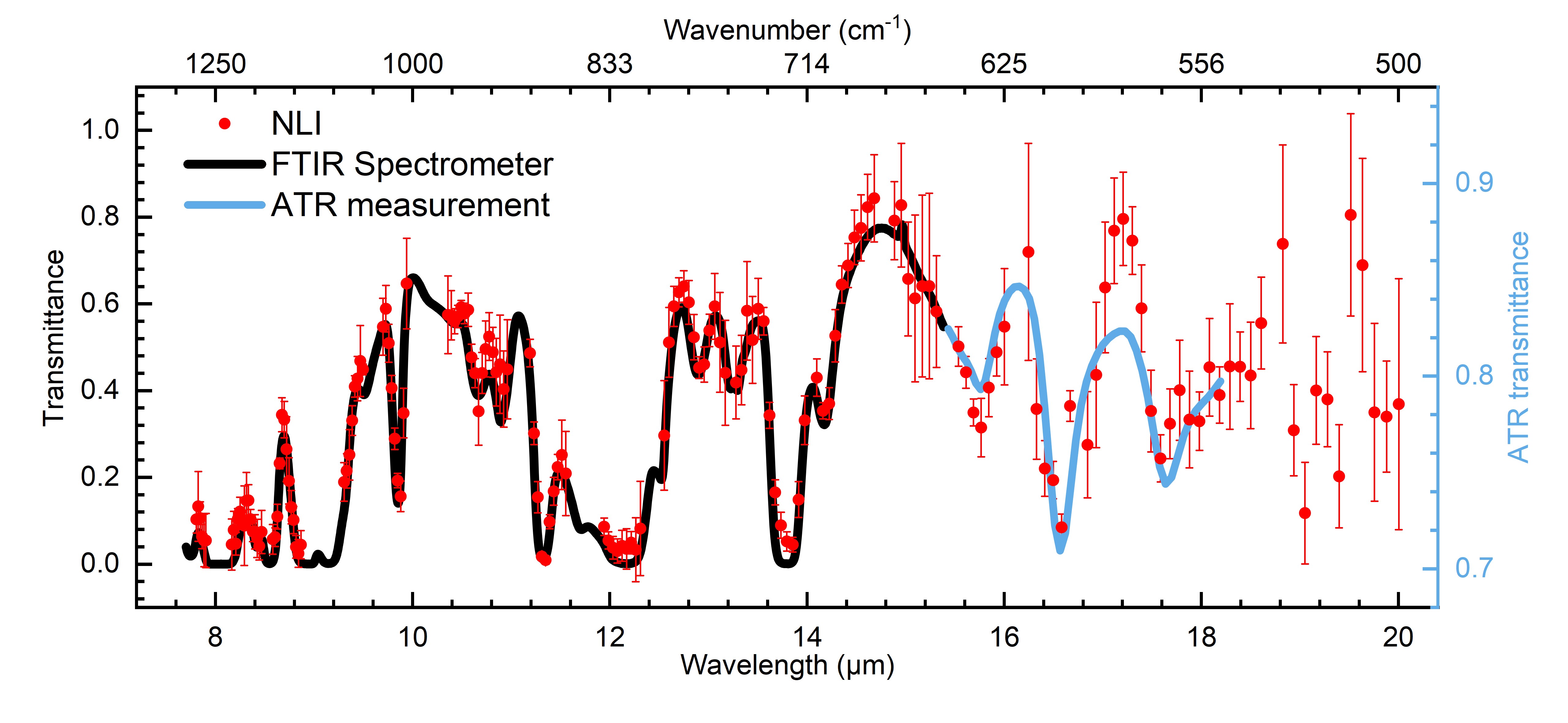}
    \caption{\justifying
    Kapton transmittance measured using the nonlinear interferometer (NLI, red points), compared with an independent FTIR transmission measurement (black curve) and an ATR measurement of the Kapton sample (blue curve). The ATR data are shown on the right axis.}
    \label{fig:kapton_spectrum}
\end{figure*}

The measurements from all values of \(\rho\) were combined to obtain the broadband Kapton transmittance spectrum shown in Fig.~\ref{fig:kapton_spectrum}. The red points show the NLI measurement, while the black curve shows an independent transmission spectrum measured using a Bruker INVENIO R FTIR spectrometer. In the available FTIR configuration, the maximum measurable wavelength was limited to \(15.385~\mu\mathrm{m}\) by the detector, beamsplitter, and other optical components. Within the available range, the NLI measurement reproduces the positions and overall shapes of the main Kapton absorption features. The largest discrepancies occur close to the strongest absorption bands, where the transmitted idler contribution approaches the noise floor and the extracted transmittance becomes less accurate.

An additional attenuated total reflection (ATR) measurement was used to provide an independent comparison at longer wavelengths. The Kapton sample was pressed against the ATR crystal and measured using a liquid-nitrogen-cooled detector. Since ATR probes only the near-surface region through the evanescent field, its transmittance magnitude is not directly comparable with that of the transmission measurement and is therefore shown on a separate right-hand axis in Fig.~\ref{fig:kapton_spectrum}. According to the instrument specification, reliable ATR data extends to approximately \(550~\mathrm{cm}^{-1}\), corresponding to \(18.2~\mu\mathrm{m}\) wavelength. Within the overlapping wavelength range, the absorption minima measured by the NLI agree well with those observed in the ATR spectrum. 

Together, the transmission FTIR and ATR measurements provide independent validation of the spectral feature positions over complementary wavelength ranges up to approximately \(18.2~\mu\mathrm{m}\), while the NLI measurement extends further to \(20~\mu\mathrm{m}\).

Although no independent reference was available over this final wavelength interval, the agreement with the FTIR and ATR measurements over their respective overlapping ranges supports the reliability of the broadband spectral measurement. The larger uncertainty towards \(20~\mu\mathrm{m}\) is consistent with the reduced SNR and SBR of the corresponding idler spectra shown in Fig.~\ref{fig:angle_spectra}.

These measurements demonstrate broadband absorption spectroscopy across \(7.8\text{--}20~\mu\mathrm{m}\) while detecting only the corresponding NIR signal photons from 817--872~nm with a silicon detector. The accessible spectroscopic range is therefore not limited by the spectral response of the detector, but primarily by the nonlinear generation and optical transmission in the undetected idler path.

\subsection{SF$_6$ gas spectroscopy}

\begin{figure}[t!]
    \centering
    \includegraphics[width=\linewidth]{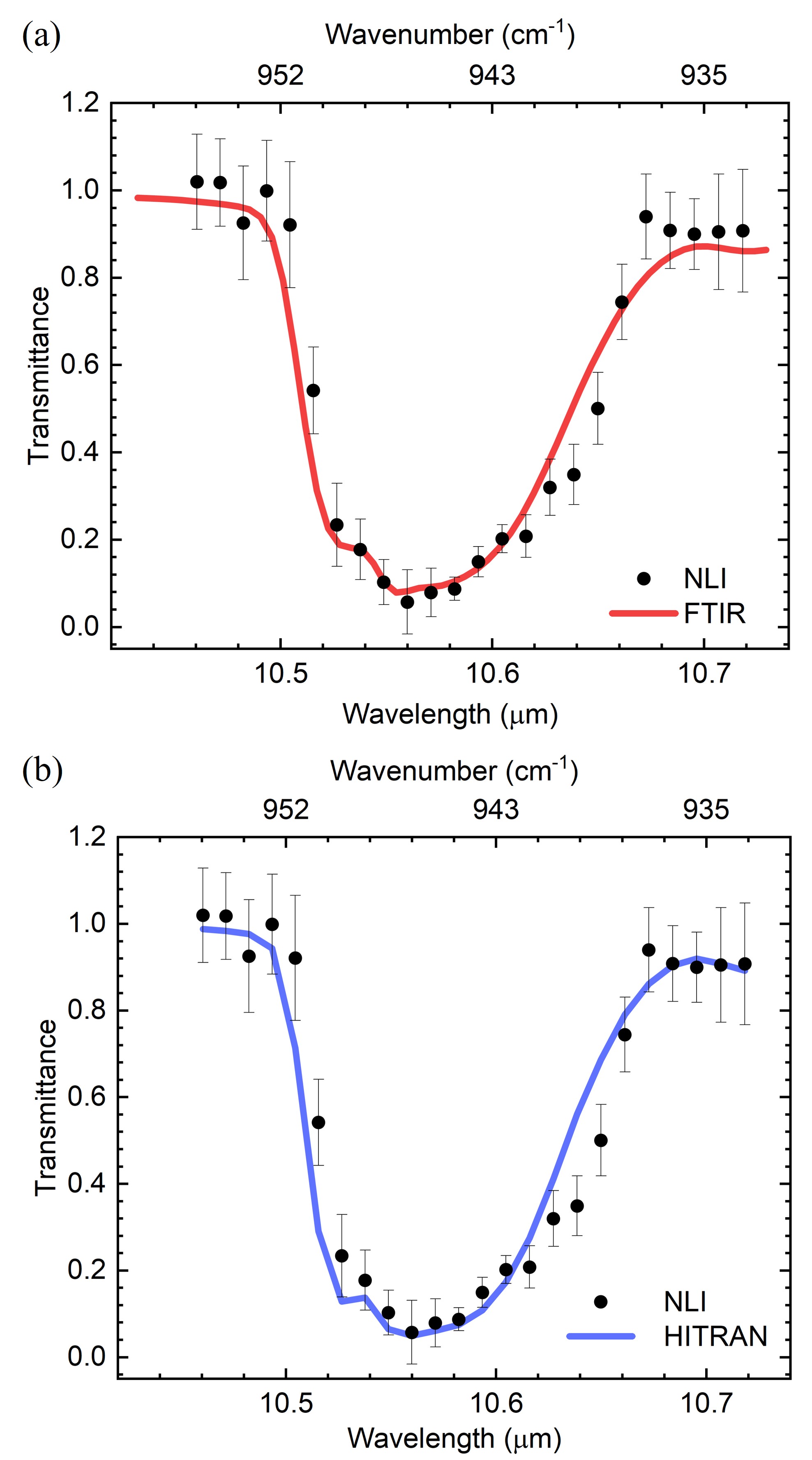}
    \caption{\justifying
    SF$_6$ gas spectroscopy using the BGGSe nonlinear interferometer.
    (a) Normalized SF$_6$ transmittance measured using the NLI (black points) and an independent FTIR spectrometer (red curve). The nominal spectral resolutions were \(1~\mathrm{cm}^{-1}\) for both NLI and FTIR measurements.
    (b) The same NLI measurements compared with the best-fit HITRAN-based SF$_6$ transmission spectrum (blue curve). The fit gives an SF$_6$ concentration of \(1132.0 \pm 58.8~\mathrm{ppm}\) for the \(5~\mathrm{cm}\) gas cell. Error bars show the uncertainty in the measured NLI transmittance.}
    \label{fig:sf6_spectrum}
\end{figure}

The gas-sensing capability of the nonlinear interferometer was demonstrated using sulfur hexafluoride (SF$_6$), which has a strong absorption band near \(10.55~\mu\mathrm{m}\)~\cite{gordon2022hitran2020}. Figure~\ref{fig:sf6_spectrum} shows the measured NLI transmittance across this absorption band.

To enable comparison of the absorption profiles, the NLI and FTIR spectra were normalized independently to their respective off-resonance transmission levels. The normalization procedure is described in the Materials and Methods section.

Representative reference and gas-cell interferograms and their corresponding Fourier spectra are shown in Supplementary Fig. S4, illustrating the change in the detected interferogram and Fourier spectrum after introduction of SF$_6$ into the idler path.

Figure~\ref{fig:sf6_spectrum}(a) compares the NLI transmittance with an independent FTIR measurement. Both measurements resolve the strong SF$_6$ absorption band centred near \(10.55~\mu\mathrm{m}\). Good agreement is observed in the position, depth, and overall width of the main absorption feature. Differences remain in the detailed spectral profile, particularly near the edges of the absorption band and in the higher-transmittance regions. These differences are consistent with the experimental uncertainty and with the different nominal spectral resolutions of the NLI and FTIR measurements.

To obtain a quantitative concentration estimate, the measured NLI spectrum was fitted using an SF$_6$ absorption model derived from HITRAN absorption cross-section data~\cite{gordon2022hitran2020}. The calculated spectrum was matched to the \(1~\mathrm{cm}^{-1}\) spectral resolution of the NLI measurement before comparison with the experimental data, as described in Materials and Methods. The best-fit spectrum is shown in Fig.~\ref{fig:sf6_spectrum}(b).

The weighted fit gives an SF$_6$ concentration of \(1132.0\pm58.8~\mathrm{ppm}\) for the \(5~\mathrm{cm}\) gas cell, corresponding to a path-integrated concentration of \(56.6\pm2.9~\mathrm{ppm\,m}\). The fitted HITRAN spectrum reproduces the position and overall profile of the measured absorption band, including the absorption minimum near \(10.55~\mu\mathrm{m}\) and the recovery in transmission on both sides of the band. As an independent quantitative comparison, the FTIR spectrum was also fitted using the HITRAN-based model, giving an SF$_6$ concentration of \(1172.0\pm7.7~\mathrm{ppm}\), consistent with the value retrieved from the NLI measurement. The FTIR concentration fit is shown in Supplementary Fig.~S6.

The agreement between the NLI and FTIR spectra, together with the consistent concentrations retrieved independently from the two measurements, demonstrates quantitative SF$_6$ spectroscopy through detection of the corresponding NIR signal field without direct detection of the infrared idler. Together with the broadband Kapton measurements, these results demonstrate both broadband material spectroscopy and molecular gas sensing using the same BGGSe nonlinear interferometer and silicon-based detection.

\section{Discussion}

The main advance of this work is the extension of Fourier-transform spectroscopy with undetected photons into a substantially longer wavelength range than previously demonstrated. Using a single BGGSe nonlinear crystal, Fourier-resolved idler spectra were obtained across a \(7.8\text{--}20~\mu\mathrm{m}\) wavelength range, while detection remained entirely in the NIR using a silicon detector. This significantly extends the accessible wavelength range of undetected-photon spectroscopy while retaining the advantages of mature, low-noise silicon detection.

The broad spectral coverage results from the wide phase-matching tunability of BGGSe. After establishing a fixed oblique-incidence geometry, rotation of the crystal provided continuous tuning of the idler spectrum while largely maintaining the interferometer alignment. Nonlinear interference was observed across widely separated idler wavelengths, and the approximately linear dependence of the idler centre wavenumber on crystal rotation provides a simple experimental means of selecting different spectral regions with the same nonlinear crystal.

The spectral magnitude, SNR, and SBR decrease towards the longer and shorter wavelengths. Several factors are likely to contribute, including wavelength-dependent nonlinear conversion efficiency, transmission through the BGGSe crystal, AR-coating losses away from their design wavelength near \(10.5~\mu\mathrm{m}\), and reduced spatial overlap of the returning idler mode at the crystal. Although the commonly reported transmission range of BGGSe extends to approximately \(18~\mu\mathrm{m}\), the observation of Fourier-resolved interference approaching \(20~\mu\mathrm{m}\) indicates that sufficient transmission remains over the short propagation length used here to support nonlinear-interferometric detection. Despite these limitations, Fourier-resolved idler peaks remained distinguishable from the background throughout the measured range, with the SNR remaining above \(9.26\). The long-wavelength performance therefore appears to be limited mainly by reduced signal quality rather than by a fundamental loss of nonlinear interference, suggesting that improved long-wavelength optics and mode matching could further extend the usable range.

The Kapton measurements provide an important broadband validation of the spectroscopic performance. Within the wavelength range covered by the transmission FTIR measurement, the absorption features measured by the NLI agree well with the FTIR spectrum. The additional ATR measurement extends this comparison to approximately \(18.2~\mu\mathrm{m}\), and the positions of the absorption features remain consistent with those obtained using the NLI. The agreement with two independent infrared measurements over their respective overlapping ranges supports the reliability of the absorption features recovered by the NLI. The NLI measurement extends further to \(20~\mu\mathrm{m}\), beyond the available range of either reference measurement, although with increased uncertainty associated with the lower SNR and SBR at the long-wavelength edge.

The \(\mathrm{SF_6}\) measurement provides a complementary demonstration of gas spectroscopy using the same platform. The absorption band near \(10.55~\mu\mathrm{m}\) was resolved using NIR signal detection and showed a spectral profile consistent with both an independent FTIR measurement and the HITRAN calculation. A weighted fit to the HITRAN-based spectrum gave a path-integrated \(\mathrm{SF_6}\) concentration of \(56.6 \pm 2.9~\mathrm{ppm\,m}\). Together with the broadband Kapton measurement, this demonstrates that the same nonlinear interferometer can be used for both broadband material spectroscopy and gas spectroscopy without direct detection of the long-wavelength idler.

The present approach is complementary to conventional FTIR spectroscopy rather than a replacement for it. Conventional FTIR remains highly effective when suitable long-wavelength sources, optics, and detectors are available. Nonlinear interferometric spectroscopy instead allows the sample to be probed at long infrared wavelengths while the measurement is performed with a shorter-wavelength detector. This separation becomes particularly attractive towards longer wavelengths, where high-performance direct detection often requires specialised detector technologies and cooling. The present long-wavelength limit is primarily set by the nonlinear generation efficiency and optical loss in the undetected idler path, rather than by the detection wavelength. Improving the long-wavelength transmission and mode overlap could extend the accessible spectral range beyond \(20~\mu\mathrm{m}\).

\section{Materials and Methods}

\subsection{BGGSe phase-matching design}

In this work, the pump wavelength was \(785~\mathrm{nm}\), while the main absorption band of \(\mathrm{SF_6}\) is near \(10.55~\mu\mathrm{m}\). The corresponding signal wavelength was determined from energy conservation,
\begin{equation}
\frac{1}{\lambda_p}
=
\frac{1}{\lambda_s}
+
\frac{1}{\lambda_i},
\label{eq:energy_conservation}
\end{equation}
where \(\lambda_p\), \(\lambda_s\), and \(\lambda_i\) are the pump, signal, and idler wavelengths. For \(\lambda_p=785~\mathrm{nm}\) and \(\lambda_i=10.55~\mu\mathrm{m}\), the signal wavelength is \(848.1~\mathrm{nm}\), which is within the detection range of silicon avalanche photodiodes.

The BGGSe crystal cut was chosen using a collinear phase-matching calculation based on its Sellmeier equations~\cite{grechin2020baga2ges6,badikov2016crystal}. Around the target wavelength pair, the calculation gave two nearby phase-matchable solutions. The \(o \rightarrow e_s+e_i\) solution was found at \(\theta=30.9^\circ\), while the \(o \rightarrow e_s+o_i\) solution was found at \(\theta=32.0^\circ\).Here \(\theta\) is the internal angle between the pump wave vector and the optical axis of the crystal. Their calculated maximum effective nonlinear coefficients were \(22.07~\mathrm{pm/V}\) and \(35.13~\mathrm{pm/V}\), respectively. The crystal cut was chosen using the \(o \rightarrow e_s+o_i\) design, which gave the larger calculated \(d_{\mathrm{eff}}\).

This calculation was used as the design configuration rather than as a direct experimental identification of the phase-matching type. Experimentally, only the NIR signal was detected, so the idler polarisation was not independently measured. The generated signal was verified by its strong dependence on crystal orientation and pump polarisation, supporting its origin from phase-matched SPDC in the BGGSe crystal rather than from residual pump leakage, fluorescence from optics, or broadband background light.

\subsection{Nonlinear interferometer setup}

The experiment used a single-mode fibre-coupled \(785~\mathrm{nm}\) continuous-wave laser system (CCS-CW, 785LD-2-SM-FBG, AeroDIODE) as the pump. After collimation, the pump passed through an optical isolator to reduce back-reflection into the laser. A half-wave plate was used to control the pump polarisation, and an \(800~\mathrm{nm}\) short-pass filter suppressed weak longer-wavelength emission from the pump source.

The pump was focused into a \(2~\mathrm{mm}\)-thick BGGSe crystal using a 6-inch off-axis parabolic mirror (OAPM), and a 3-inch OAPM was used after the crystal to collimate the pump and generated fields. The shorter focal length helped to limit the beam size of the more strongly diverging long-wavelength idler and reduce clipping in the idler arm.

A short-pass dichroic mirror separated the idler from the pump and signal. After the second pass, the signal was separated using a long-pass dichroic mirror and spectral filters. An \(800~\mathrm{nm}\) long-pass filter suppressed residual pump light, while additional filters were selected according to the generated signal wavelength. The signal was coupled into a multimode fibre and detected using a silicon single-photon counting module (SPCM-AQR-12-FC, Excelitas Technologies).

The pump polarisation was re-optimised at each crystal rotation angle \(\rho\) using the half-wave plate to maximise the detected signal photon count rate. The available signal filters covered wavelengths from approximately \(816~\mathrm{nm}\) to \(870~\mathrm{nm}\). The collection optics sampled mainly the central region of the generated signal beam, favouring the near-collinear emission that provided the strongest nonlinear interference. Rotating the pump polarisation optimised the achievable \(d_{\mathrm{eff}}\) at each \(\rho\), ensuring sufficient signal generation was still observed at all measured values of \(\rho\), allowing the interferometric measurements to be performed.

\subsection{Interferogram acquisition and Fourier-transform analysis}

The idler-arm mirror was mounted on a motorised linear translation stage (Newport, LTA-HL). To obtain the idler spectra, the mirror was scanned over a displacement of \(1.6~\mathrm{mm}\) with a step size of \(0.1~\mu\mathrm{m}\). In the Michelson configuration, this corresponded to a total optical path difference of \(3.2~\mathrm{mm}\) and an optical path difference increment of \(0.2~\mu\mathrm{m}\), giving a Fourier-transform resolution of approximately \(3.125~\mathrm{cm}^{-1}\). The SPCM integration time at each position was \(100~\mathrm{ms}\).

For the local fringe measurement shown in Fig.~\ref{fig:fringes}, the idler-arm mirror was scanned with a step size of \(0.4~\mu\mathrm{m}\), corresponding to an optical path difference increment of \(0.8~\mu\mathrm{m}\). Residual pump and signal leakage produced higher-frequency components that were spectrally separated from the idler contribution in the Fourier domain.

To quantify the detectability of the recovered idler spectra, the signal-to-noise ratio (SNR) and signal-to-background ratio (SBR) were calculated from the FFT magnitude. For each spectrum, the idler peak was first identified within the expected \(7\text{--}20~\mu\mathrm{m}\) wavelength range. A Savitzky--Golay filter was used to locate the peak and determine the boundaries of the signal region, but all final SNR and SBR values were calculated from the original FFT data.

The background was evaluated within a \(10~\mu\mathrm{m}\)-wide spectral window centred on the identified idler peak. The signal region was excluded together with an additional \(0.3~\mu\mathrm{m}\) guard interval on each side to reduce contamination of the background estimate by the spectral peak. The mean background FFT magnitude, \(\overline{A}_{\mathrm{bg}}\), and its standard deviation, \(\sigma_{\mathrm{bg}}\), were then calculated from the remaining points. The SNR was defined as
\begin{equation}
\mathrm{SNR}
=
\frac{
A_{\mathrm{peak}}-\overline{A}_{\mathrm{bg}}
}{
\sigma_{\mathrm{bg}}
},
\label{eq:snr}
\end{equation}
where \(A_{\mathrm{peak}}\) is the maximum FFT magnitude within the identified signal region. The SBR was defined as
\begin{equation}
\mathrm{SBR}
=
\frac{
A_{\mathrm{peak}}
}{
\overline{A}_{\mathrm{bg}}
}.
\label{eq:sbr}
\end{equation}

The photon-count interferogram recorded at each crystal angle was Fourier-transformed to obtain the corresponding idler spectrum. For sample measurements, reference and sample interferograms were recorded under the same experimental conditions. The transmittance at each idler wavenumber was calculated from the ratio of the corresponding Fourier amplitudes,
\begin{equation}
T(\tilde{\nu})
=
\frac{
A_{\mathrm{sam}}(\tilde{\nu})
}{
A_{\mathrm{ref}}(\tilde{\nu})
},
\label{eq:nli_transmittance_methods}
\end{equation}
where \(A_{\mathrm{sam}}\) and \(A_{\mathrm{ref}}\) are the Fourier amplitudes measured with and without the sample, respectively. In the induced-coherence interferometer, the signal-interference amplitude is proportional to the field transmission of the idler between the two nonlinear interactions~\cite{PhysRevA.44.4614}. In the present Michelson geometry, the idler passes through the sample twice. For a single-pass intensity transmittance \(T\), the field transmission is \(\sqrt{T}\) per pass, giving a double-pass field transmission of \(T\). The Fourier-amplitude ratio therefore directly yields the single-pass intensity transmittance of the sample. The corresponding idler wavelength was calculated as \(\lambda_i=10^4/\tilde{\nu}\), where \(\lambda_i\) is in \(\mu\mathrm{m}\) and \(\tilde{\nu}\) is in \(\mathrm{cm}^{-1}\).

\subsection{Kapton transmittance measurement}

For each crystal angle, a reference scan was first recorded without the sample in the idler path. The Kapton sample was then inserted into the idler arm, and the measurement was repeated at the same crystal angle. The scan range, step size, and SPCM integration time were kept the same between the reference and sample measurements. The Kapton transmittance was calculated using Eq.~\ref{eq:nli_transmittance_methods}, and measurements obtained at different crystal angles were combined to obtain the broadband transmittance spectrum.

Two independent FTIR measurements were performed for comparison. A transmission spectrum was measured using a Bruker INVENIO R FTIR spectrometer in the available transmission configuration. An additional measurement was performed using a ConcentratIR2 ATR accessory with a liquid-nitrogen-cooled detector. For the ATR measurement, the Kapton sample was pressed against the ATR crystal to ensure good optical contact. In contrast to transmission spectroscopy, the ATR measurement probes the sample through the evanescent field generated at the crystal--sample interface and is therefore primarily sensitive to a shallow region near the sample surface rather than the full Kapton thickness. The diamond ATR configuration has a specified lower wavenumber limit of approximately \(550~\mathrm{cm}^{-1}\), corresponding to \(18.2~\mu\mathrm{m}\). The ATR spectrum reported as transmittance by the instrument software was used for comparison of the spectral feature positions with the NLI measurement, rather than for a direct comparison of absolute transmittance.

\subsection{SF$_6$ gas-cell measurement and concentration retrieval}

For the SF$_6$ measurement, the crystal angle was selected so that the generated idler spectrum overlapped the strong absorption band near \(10.55~\mu\mathrm{m}\). A reference interferogram was first recorded without the gas cell in the idler path. The SF$_6$ gas cell, with an optical path length of \(L=5~\mathrm{cm}\), was then inserted into the idler arm. The cell was fitted with two \(2~\mathrm{mm}\)-thick AR-coated ZnSe windows.

Insertion of the ZnSe windows increased the optical path length of the idler arm and shifted the coherence position of the nonlinear interferometer. The signal-arm mirror was initially adjusted according to the refractive index and total thickness of the ZnSe windows to compensate for this additional optical path. The coherence maximum was subsequently located experimentally for both the reference and gas-cell measurements. The interferograms were then analysed over the same optical path difference range centred on their respective coherence maxima, ensuring that equivalent regions of the two interferograms were used for the Fourier analysis.

For the SF$_6$ measurement, the idler-arm mirror was scanned over a total displacement of \(5~\mathrm{mm}\). In the Michelson configuration, this corresponds to a total optical path difference of \(10~\mathrm{mm}\), giving a nominal Fourier-transform spectral resolution of \(1~\mathrm{cm}^{-1}\). The reference and gas-cell interferograms were Fourier-transformed separately, and the wavelength-dependent raw transmission was calculated from the ratio of their Fourier amplitudes using Eq.~\ref{eq:nli_transmittance_methods}.

An off-resonance region from \(10.46\) to \(10.49~\mu\mathrm{m}\), where SF$_6$ absorption is negligible, was used to determine the baseline transmission. The mean raw transmittance, obtained from the Fourier-amplitude ratio over all data points within this interval, was \(0.80\). Since no appreciable SF$_6$ absorption is expected in this region, the reduction from unity was attributed primarily to wavelength-independent insertion losses associated with introducing the gas cell into the idler path. The NLI transmission spectrum was therefore normalized by this mean off-resonance value, so that the baseline transmittance was set to unity,
\begin{equation}
T_{\mathrm{NLI}}(\tilde{\nu})
=
\frac{
T_{\mathrm{raw}}(\tilde{\nu})
}{
0.80
}.
\label{eq:sf6_normalisation}
\end{equation}
The corresponding transmittance uncertainties were divided by the same factor.

The independent FTIR measurement had an off-resonance transmittance of approximately \(0.41\) and was likewise normalized to unity over its off-resonance region. The FTIR spectrum was recorded at a spectral resolution of \(1~\mathrm{cm}^{-1}\). This normalization places the NLI and FTIR measurements on the same relative-transmittance scale as the HITRAN model, enabling direct comparison of their spectral profiles.

Quantitative concentration retrieval was performed using the SF$_6$ absorption cross-section dataset from HITRAN~\cite{gordon2022hitran2020}. The reported spectral resolution of the cross-section dataset was approximately \(0.112~\mathrm{cm}^{-1}\), and the tabulated absorption cross section, \(\sigma(\tilde{\nu})\), was expressed in units of \(\mathrm{cm^2\,molecule^{-1}}\).

For a trial SF$_6$ concentration \(C\), expressed in ppm, the high-resolution transmission spectrum was calculated from the Beer--Lambert law,
\begin{equation}
T_{\mathrm{HR}}(\tilde{\nu};C)
=
\exp
\left[
-\sigma(\tilde{\nu})
N
\left(C\times10^{-6}\right)
L
\right],
\label{eq:sf6_beer_lambert}
\end{equation}
where \(N\) is the total gas number density in molecules per cubic centimetre and \(L=5~\mathrm{cm}\) is the gas-cell path length. The gas number density was calculated from the ideal-gas relation,
\begin{equation}
N
=
\frac{P}{k_{\mathrm{B}}T},
\label{eq:sf6_number_density}
\end{equation}
with the appropriate conversion from molecules per cubic metre to molecules per cubic centimetre.

To account for the finite spectral resolution of the NLI measurement, the high-resolution transmission spectrum calculated from Eq.~\ref{eq:sf6_beer_lambert} was convolved with a Gaussian instrument response having a full width at half maximum of \(1~\mathrm{cm}^{-1}\). The convolved model spectrum was then interpolated at the experimentally measured NLI wavenumbers. The convolution was applied to the calculated transmission spectrum rather than directly to the absorption cross section.

The SF$_6$ concentration was obtained using a weighted nonlinear least-squares fit, with the concentration \(C\) as the only free parameter. The fit minimized
\begin{equation}
\chi^2(C)
=
\sum_i
\left[
\frac{
T_i
-
T_{\mathrm{model}}(\tilde{\nu}_i;C)
}{
\delta T_i
}
\right]^2,
\label{eq:sf6_chi_square}
\end{equation}
where \(T_i\) and \(\delta T_i\) are the calibrated NLI transmittance and its uncertainty at wavenumber \(\tilde{\nu}_i\), respectively, and \(T_{\mathrm{model}}\) is the HITRAN-based transmission spectrum after convolution to the experimental resolution. The best-fit concentration was \(1132.0~\mathrm{ppm}\), with a statistical fitting uncertainty of \(58.8~\mathrm{ppm}\).

\begin{acknowledgments}
The authors acknowledge support from EPSRC through the QuSIT Hub (EP/Z533166/1), and the China Scholarship Council (CSC)/University of Bristol joint-funded scholarship program (CSC 02208410046). ASC acknowledges support from The Royal Society (URF/R/221019, RF/ERE/210098, RF/ERE/221060).
\end{acknowledgments}

\section*{Author details}

\textsuperscript{1}Quantum Engineering Technology Labs, H.H. Wills Physics Laboratory and School of Electrical, Electronic and Mechanical Engineering, University of Bristol, Bristol BS8 1TL, UK.

\section*{Author contributions}

J.D., W.N., J.G.R and A.S.C. conceived the study and designed the experiment. J.D. performed the experiments, with J.Z. contributing to data acquisition. J.D and H.Z. developed the hardware-control code. J.D. analysed the data and wrote the manuscript. A.S.C. and J.G.R. supervised the project and contributed to manuscript revision.

\section*{Data Availability Statement}
The data that support the findings of this study are available from the corresponding author upon reasonable request.

\section*{Conflict of Interest}
The authors declare no conflict of interest.

\renewcommand{\bibsection}{\section*{References}}
\bibliography{aipsamp}

\end{document}